\documentclass{elsart3p}

\usepackage{epsfig}
\usepackage{amssymb}
\usepackage{graphicx}
\usepackage{subfigure}
\usepackage{here}
\usepackage{hyperref}

\def\figurename{\bf{Fig.\space}}

\begin{document}

\begin{frontmatter}

\title{High-gain DC-mode operated Gaseous Photomultipliers for the visible spectral range}

\author[w]{A. Lyashenko\corauthref{cor}},
\corauth[cor]{Corresponding author: tel. +972-8-934-2064, fax
+972-8-934-2611, E-mail: alexey.lyashenko@weizmann.ac.il}
\author[w]{A. Breskin},
\author[w]{R. Chechik},
\author[p1]{F.D. Amaro},
\author[p1,p2]{J.F.C.A. Veloso} and
\author[p1]{J.M.F. dos Santos}
\address[w]{Department of Particle Physics, The Weizmann Institute of Science, 76100 Rehovot, Israel}
\address[p1]{Physics Department, University of Coimbra, 3004-516 Coimbra,  Portugal}
\address[p2]{Physics Department, University of Aveiro, Campus Universit\'ario de Santiago, 3810-193 Aveiro, Portugal}

\begin{abstract}

We shortly describe recent progress in photon detectors combining
bi-alkali photocathodes and cascaded patterned gas-avalanche
electron multipliers. It permitted the development and the first
feasibility demonstration of high-gain gaseous photomultipliers
sensitive in the visible spectral range, operated in DC mode with
single-photon sensitivity.

\end{abstract}

\end{frontmatter}

\section{Introduction}

Gaseous photomultipliers (GPM) have been widely employed since a few
decades for single-photon imaging in the UV spectral range - mainly
in Cherenkov detectors. Their main advantage is the possibility of
conceiving large-area atmospheric-pressure devices, with low
sensitivity to magnetic fields, having multiplication factors that
permit efficient imaging of light at single-photon levels. In recent
years there has been a considerable progress in the development GPM
concepts with advanced Micropattern electron multipliers, as
reviewed in \cite{chechik:08}. In most cases these were coupled to
CsI UV-sensitive photocathodes. Intensive R\&D efforts have been
motivated by the necessity to overcome some basic limitations of
wire-chamber multipliers; others were directed towards the possible
extension of GPMs' sensitivity from the UV to the visible spectral
range. Here the main difficulties are the photocathodes' (e.g.
bi-alkali) chemical reactivity, as well as gain limitations due to
"feedback effects" caused by photon- and ion-mediated
secondary-avalanches. The very high reactivity of visible-sensitive
photocathodes results in very short lifetimes in gases, even with
impurity levels in the fraction of ppm range. Therefore,
visible-sensitive GPMs can operate only in sealed containers, as
shown in \cite{balcerzyk:03}. Efforts were made to coat bi-alkali
photocathodes with thin protective films \cite{peskov:94,shefer:02},
as to allow for gas-flow operation; these resulted, at best, in
residual QE-values of 4-5 fold lower than that of bare photocathodes
\cite{shefer:02,shefer:thes}. Feedback effects limit the detector's
gain and its single-photon detection efficiency; they affect time
and position resolutions and damage the photocathode. The
ion-induced secondary-electron emission is particularly important in
GPMs with visible-sensitive photocathodes; their low
electron-emission threshold clearly limits the avalanche gain in DC
mode \cite{balcerzyk:03}. Visible-sensitive GPMs reached however
high gains ($\sim10^6$) with pulsed ion-gating in cascaded GEMs
\cite{breskin:05}. The present success in their stable DC operation,
at gains of $10^5$, resulted from an efficient avalanche-ion
blocking with advanced patterned hole-multipliers
\cite{lyashenko:06,lyashenko:07}. The various approaches of
multiplier-concepts for visible-sensitive GPMs are reviewed in
\cite{chechik:08}. More information about sealed visible-sensitive
GPMs, bi-alkali photocathode production and high-gain gated
operation is provided in \cite{moermann:thes}.

\section{Visible-sensitive GPM with cascaded patterned hole-multipliers}

In cascaded gaseous "hole-multipliers" of different structures,
discussed in \cite{chechik:08}, the avalanche develops in successive
multiplication stages and is confined within the holes. Cascading
several successive multipliers results in total avalanche gains
above $10^5$ - adequate for efficient single-photon detection. Most
of the avalanche-induced secondary photons originating from the last
avalanche in the cascade are efficiently screened by the cascade
elements, thus preventing photon-feedback effects. Ion-feedback
reduction is by far more difficult and challenging. It is inherently
difficult to prevent avalanche ions from back-drifting to the
photocathode while maintaining high multiplier's gain and full
photoelectron collection and detection efficiencies; this is because
the ions follow the same field lines (though in an opposite
direction) as the avalanche electrons. Efficient methods were
recently developed that permit a very significant reduction of the
Ion Back-flow Fraction (IBF), e.g. the fraction of total
avalanche-generated ions reaching the photocathode in a GPM
\cite{lyashenko:06,lyashenko:07}.

\begin{figure}[!h]
\begin{center}
\renewcommand{\thesubfigure}{\thefigure\alph{subfigure}}
\makeatletter
\renewcommand{\@thesubfigure}{\Large(\alph{subfigure})}
\renewcommand{\p@subfigure}{Fig.\space}
\renewcommand{\p@figure}{Fig.\space}
\makeatother %
\subfigure
{
    \label{fig:GEM}
    \includegraphics[width=3.5cm]{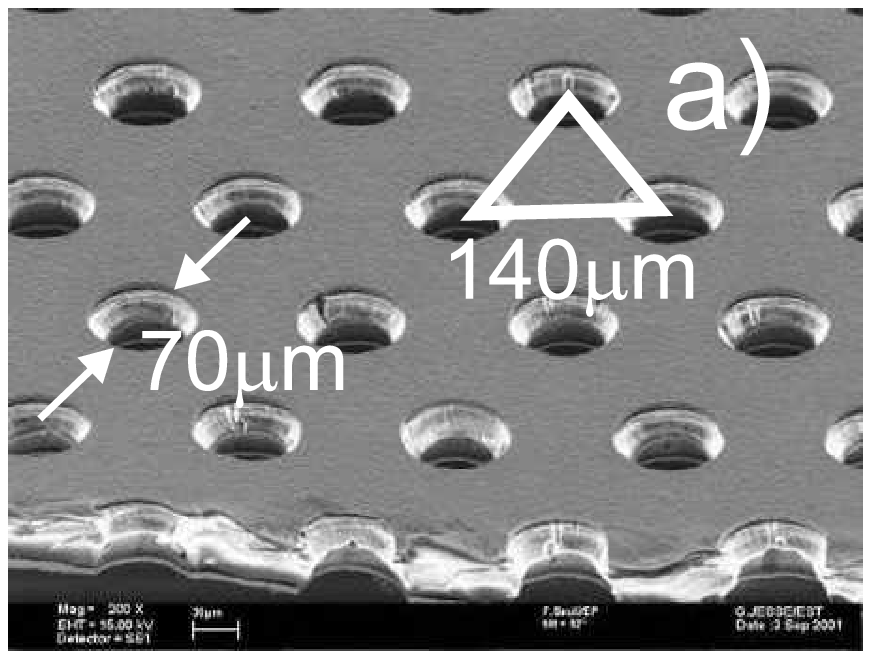}
}\hspace{0.1cm} %
\subfigure
{
    \label{fig:MHSP}
    \includegraphics[width=3.5cm]{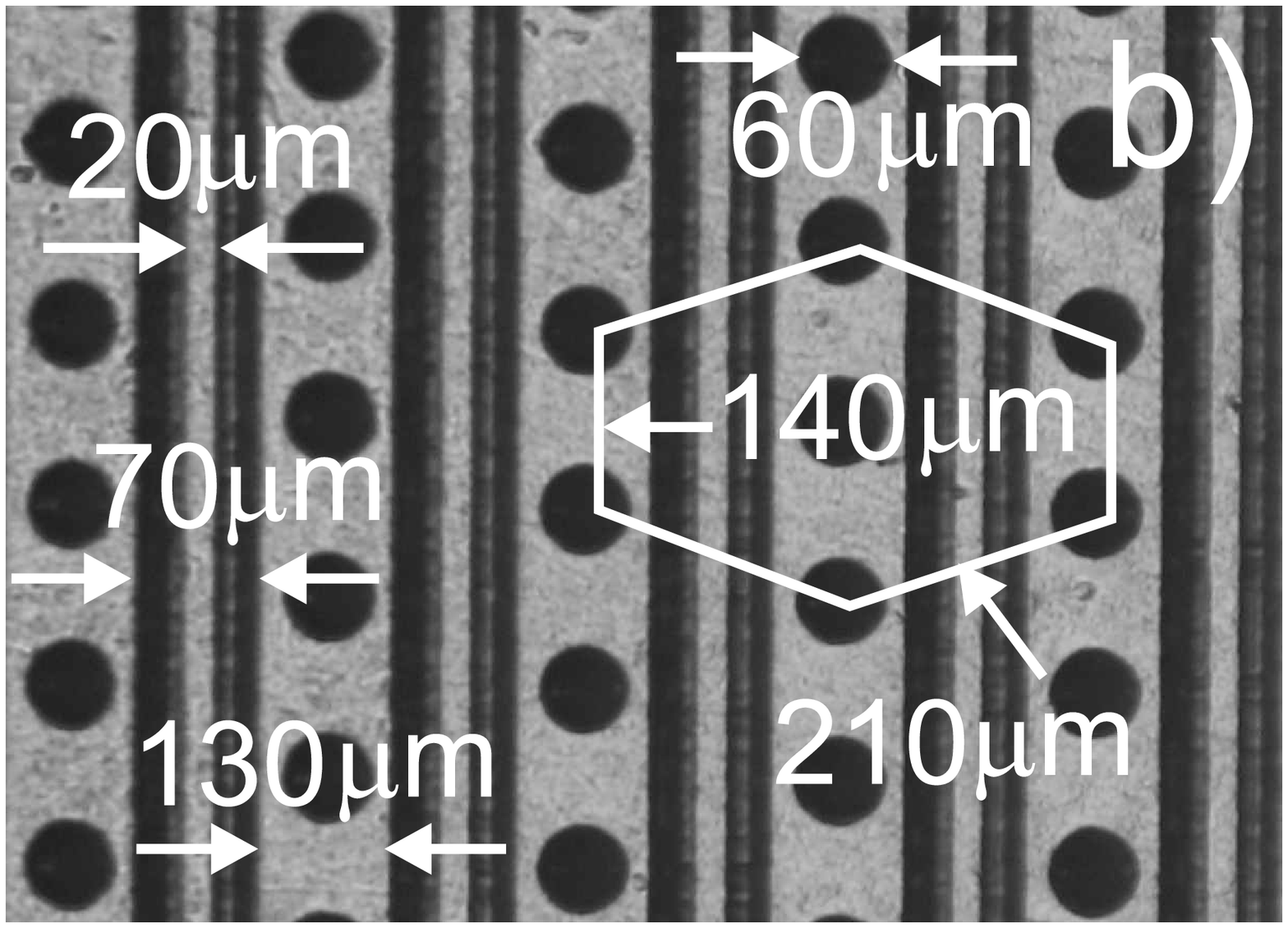}
}%
\caption {Photographs of gaseous electron hole-multipliers used in
this work. a) the GEM, with two identical faces and b) the MHSP
strip-patterned face; its other face is GEM-like.}
\label{fig:micro}
\end{center}
\end{figure}

The lowest ever attained IBF-values, of $3\cdot10^{-4}$, were
reached, at gains of $10^5$ and with full single-photoelectron
collection efficiency \cite{lyashenko:07} combining Gaseous Electron
Multipliers (GEM) \cite{sauli:97} and other patterned
hole-multipliers: Micro Hole \& Strip Plates (MHSPs)
\cite{veloso:00} (\ref{fig:GEM},\ref{fig:MHSP}). Both have
micro-holes (typically 50-70 microns in diameter) densely etched in
a thin double-sided metal-clad 50 microns thick polyimide. In GEM,
multiplication occurs under high electric fields within the holes.
The MHSP has thin anode- and cathode-strips patterned on the
GEM-like electrode; the anode strips either multiply electrons
(following the initial hole-multiplication) or, with reversed strip
polarity, trap ions - preventing them from drifting backwards to the
photocathode \cite{lyashenko:06,lyashenko:07}. In the ion-defocusing
mode, the MHSP strips can either point towards the successive
cascade elements, in reversed-bias MHSP (R-MHSP) mode, trapping
their ions, or to point towards the photocathode; in the latter, the
so-called flipped reverse-bias MHSP (F-R-MHSP), the strips also trap
the avalanche ions generated within this multiplier's holes. In both
cases conditions were found for efficient ion blocking under full
photoelectron collection efficiency
\cite{lyashenko:06,lyashenko:07}. Based on the recently measured
ion-induced electron emission probabilities from bi-alkali
photocathodes, the low IBF value of $3\cdot10^{-4}$ fulfills the
requirement for stable DC operation of visible-sensitive GPMs at
gains of $10^5$ \cite{chechik:08}.

\begin{figure} [h]
  \begin{center}
  \makeatletter
    \renewcommand{\p@figure}{Fig.\space}
  \makeatother
    \epsfig{file=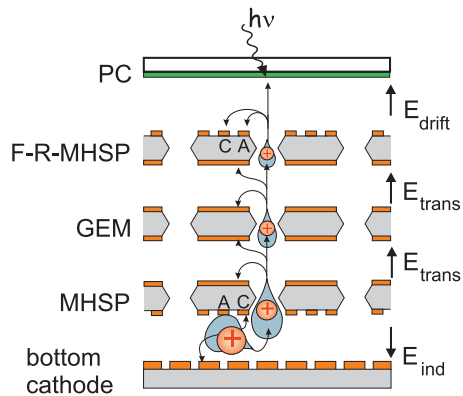, width=6cm}
    \caption{The gaseous visible-sensitive photomultiplier (GPM) comprising: a semitransparent bi-alkali photocathode followed by 3 different cascaded electron multipliers: F-R-MHSP, GEM and MHSP (with 4 avalanche stages). The arrows show possible avalanche-ions paths.}
    \label{fig:scheme}
  \end{center}
\end{figure}

\begin{figure} [h]
  \begin{center}
  \makeatletter
    \renewcommand{\p@figure}{Fig.\space}
  \makeatother
    \epsfig{file=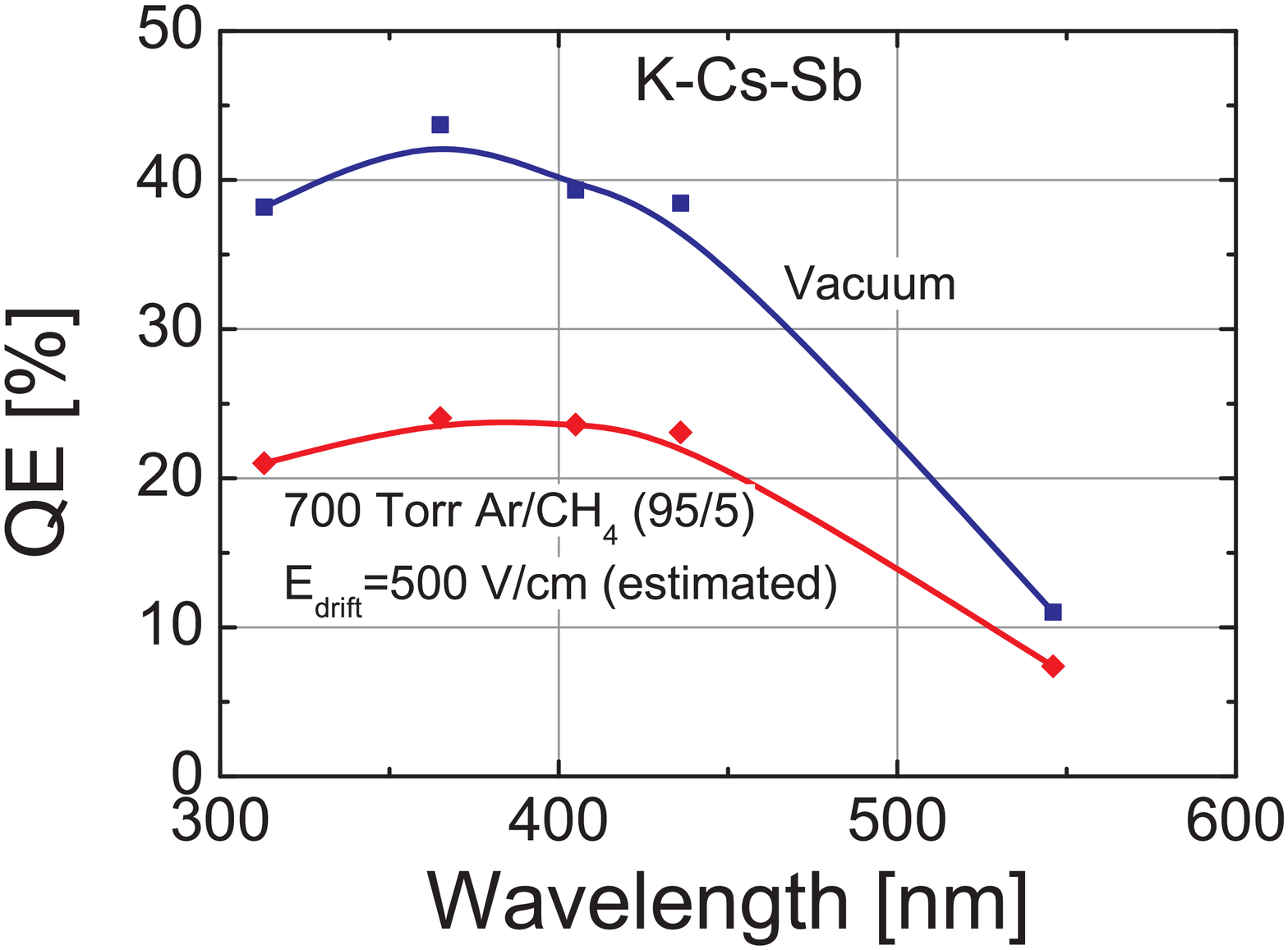, width=8cm}
    \caption{Typical quantum efficiency curve - measured in vacuum after production and the one in Ar/CH$_4$(95/5) at 700
Torr; the latter was estimated on the basis of photoelectron
backscattering under 0.5 kV/cm  \protect\cite{breskin:02}.}
    \label{fig:QE}
  \end{center}
\end{figure}

Following this successful ion blocking, a first proof of principle
was recently made in the DC-mode operation of a visible-sensitive
GPM. The detector comprised a semitransparent K-Cs-Sb bi-alkali
photocathode coupled to a cascaded hole-multiplier composed of a
F-R-MHSP followed by a GEM and an MHSP, described above
(\ref{fig:scheme}). The QE of this GPM in Ar/CH$_4$(95/5) at 700
Torr was estimated, on the basis of electron-backscattering data
\cite{breskin:02}, from the vacuum-QE measured after the
bi-alkali-photocathde production (\ref{fig:QE}). Values of 24\% were
estimated at 400 nm.

\begin{figure} [h]
  \begin{center}
  \makeatletter
    \renewcommand{\p@figure}{Fig.\space}
  \makeatother
    \epsfig{file=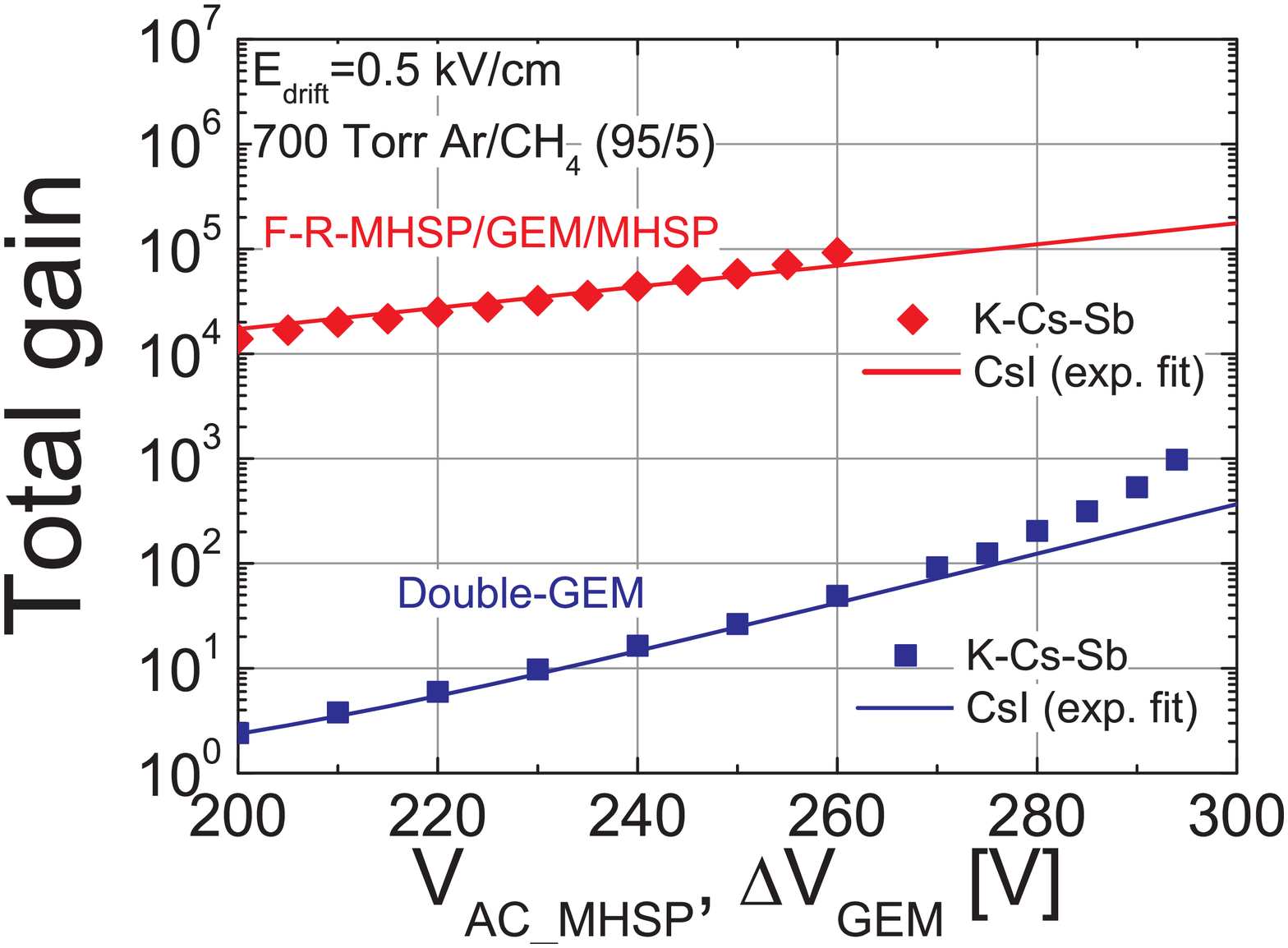, width=8cm}
    \caption{Gain curves measured in the GPM of \protect\ref{fig:scheme} and, for comparison, in a GPM of 2 cascaded GEMs coupled to a semitransparent photocathode. In addition to the curves with bi-alkali photocathodes (data points), are shown lines fitted to the data measured with CsI UV-photocathodes in the same conditions. Notice the gain "divergence" in the bi-alkali/double-GEM GPM already at gains $>$100.}
    \label{fig:K-Cs-Sb_gain}
  \end{center}
\end{figure}

Stable operation at gains of $10^5$ (not the detector's limit) was
reached in DC mode, at 700 Torr of Ar/CH$_4$ (95/5)
(\ref{fig:K-Cs-Sb_gain}). The response of a GPM composed of 2
cascaded GEMs coupled to a bi-alkali photocathode is shown for
comparison; notice the gain "divergence" in the bi-alkali/double-GEM
GPM, occurring already at gains above 100, compared to the regular
exponential behavior curve obtained with the cascaded GPM of
\ref{fig:scheme}. This validated the hypothesis that an efficient
ion blocking (here IBF=$3\cdot10^{-4}$) permitted, for the first
time, operating a visible-sensitive gaseous photomultiplier at such
high gains. In addition to the curves with bi-alkali photocathodes
(data points), lines fitted to the data measured with CsI
UV-photocathodes are shown in \ref{fig:K-Cs-Sb_gain}, under the same
conditions; the stable operation with CsI is due to the lack of
ion-feedback. Large-area UV-sensitive cascaded-GEM GPMs with CsI
photocathodes are already in use for single-photon imaging in
Cherenkov detectors \cite{chechik:08,fraenkel:05}.

The ageing of semitransparent K-Cs-Sb photocathodes under
avalanche-ion impact was recently investigated \cite{breskin:05}. The measurements
showed typically a QE decay of $\sim20\%$ after an accumulated
charge (avalanche ions) of 1-2 $\mu$C/mm$^2$ on the photocathode. In
terms of a photon detector with a bi-alkali PC, operating at a gain
of $10^5$ and assuming an IBF value of $3\cdot10^{-4}$, a 20\% QE
drop would occur after 46 years of operation at a photon rate of 5
kHz/mm$^2$. For comparison, due to the high ion-backflow in a
MWPC-based photon detector, the photocathode would only survive 5
days under the same operation conditions \cite{lyashenko:08}.

\section{Summary}

Following an intensive and long R\&D program, we demonstrated, for
the first time, the possibility of operating gaseous
photomultipliers (GPM) sensitive in the visible spectral range, at
high gains, in DC mode. The stable operation, free of ion-feedback
gain-limitations, was reached with cascaded hole-multipliers
combining three different electron-multiplier elements: a F-R-MHSP,
a GEM and a MHSP. The 3 different multipliers permitted reaching
high gains, with 100\% photoelectron collection efficiency and with
unprecedented ion-blocking capability. The latter was reached by ion
trapping with strips patterned on the surfaces of the MHSP-like
elements. Due to the small hole-size and pitch, cascaded GEM and
GEM-like GPMs offer 2D single-photon localization resolutions in the
100 micrometer range \cite{sauli:05,maia:07}. Large-area photon
detectors could also use the more recent Thick-GEM (THGEM) cascaded
multipliers \cite{breskin:08} with patterned ion-blocking
electrodes. These economically manufactured devices have larger
holes but still sub-millimeter resolutions \cite{cortesi:07}. Such
devices would require low-outgassing substrate-materials (e.g.
polyimide, ceramic, glass etc.) when applied to visible-sensitive
GPMs. The prospects of producing economically large-area, sealed
flat visible-photon detectors of this kind, with single-photon
sensitivity and good localization and timing properties, capable of
operation at MHz/mm$^2$ rates - should be very challenging to
industry! These would pave ways towards numerous potential
applications.

\section*{Acknowledges}

The work was supported inpart by grants from the Israel Science
Foundation (Project No. 151/01), the MINERVA Foundation (project No.
708566) and the Foundation for Science \& Technology - Lisbon,
project POCI/FP/81955/2007. A. Breskin is the W.P.Reuther Professor
of Research in peaceful uses of atomic energy.

\bibliographystyle{elsart-num}
\bibliography{publications}

\end{document}